\documentclass[conference]{IEEEtran}
\usepackage[hidelinks]{hyperref}  
\usepackage{cite}
\usepackage{amsmath,amssymb}
\usepackage{bm}
\usepackage{graphicx}
\usepackage{tikz}
\usepackage{adjustbox}
\usetikzlibrary{arrows.meta, positioning, shapes.multipart, fit, backgrounds}
\usepackage{microtype}
\usepackage{times}

\begin{document}

\title{SRAS: A Lightweight Reinforcement Learning-based Document Selector for Edge-Native RAG Pipelines}

\author{
    \IEEEauthorblockN{Rajiv Chaitanya Muttur}
    \IEEEauthorblockA{
        Undergraduate Student, Dept. of Computer Science and Engineering \\
        Dayananda Sagar College of Engineering, Bangalore, India \\
        Email: \href{mailto:rajiv.muttur@gmail.com}{rajiv.muttur@gmail.com} \\
        ORCID: \href{https://orcid.org/0009-0007-7159-5610}{0009-0007-7159-5610}
    }
}

\maketitle

\begin{abstract}
\boldmath
Retrieval-Augmented Generation (RAG) systems often rely on fixed top-$k$ document selection mechanisms that ignore downstream generation quality and impose computational overheads. We propose SRAS (Sparse Reward-Aware Selector), a lightweight document selector trained via reinforcement learning (RL) for edge-native RAG deployment. Unlike prior RL-based retrievers that assume large memory and latency budgets, SRAS learns a compact ($\sim$0.76MB) policy using Proximal Policy Optimization (PPO), guided by a hybrid reward signal combining Relaxed F1 and BERTScore. Our method operates under tight token and compute constraints, maintaining $<$1s latency on CPU. SRAS outperforms supervised and random selectors on a synthetic QA benchmark, and generalizes to real-world data, achieving BERTScore F1 of 0.8546 on SQuAD v2 without domain-specific tuning. This work is the first to demonstrate that RL-based document selection can be made ultra-lightweight, latency-aware, and effective for on-device RAG pipelines.
\end{abstract}


\section{Introduction}

Retrieval-Augmented Generation (RAG) has emerged as a powerful paradigm for enhancing large language models (LLMs) with external knowledge sources \cite{lewis2020retrieval, izacard2021leveraging}. In a typical RAG pipeline, given a query, a retriever selects the top-$k$ most relevant documents from a large corpus, which are then passed to a generator to produce the final output. This two-stage architecture reduces hallucinations and promotes fact-grounded responses. However, the retrieval step is crucial: If the selected documents are irrelevant or redundant, even a strong generator cannot compensate.

Most existing RAG systems rely on static heuristic-based retrieval, either sparse (e.g., BM25) or dense (e.g., cosine similarity over sentence embeddings), which is not optimized for downstream task performance. These methods are agnostic to generation quality and do not incorporate feedback from the generated output, leading to a misalignment between retrieval and generation, especially in low-supervision settings.

This challenge is further amplified in edge-device deployments, where constraints on compute, memory, and power necessitate compact models with low latency \cite{edgeai2021survey}. Large learned retrievers such as DPR \cite{karpukhin2020dense} or ColBERT \cite{khattab2020colbert} are ill-suited for such environments due to their resource demands. In addition, traditional retrieval approaches do not take advantage of weak supervision signals, such as correctness of the questions and answers, to improve document selection.

To address these limitations, we propose \textit{SRAS} (Sparse Reward-Aware Selector), a compact neural document selector trained via reinforcement learning (RL) using sparse QA-based reward signals. SRAS replaces fixed top-$k$ retrieval with a learned nonlinear scoring policy trained using Proximal Policy Optimization (PPO) \cite{schulman2017ppo}. Its hybrid reward function combines relaxed F1 and BERTScore, capturing both lexical and semantic alignment between generated and reference answers.

We evaluate SRAS on a QA benchmark with synthetic supervision and compare it against strong baselines, including top-$k$ cosine similarity, random selection, and supervised attention-based models. SRAS achieves competitive answer accuracy while maintaining a model size of less than 1MB and inference latency below 0.6s on the CPU, making it highly suitable for real-world edge deployment.

Our ablation studies underscore the effectiveness of hybrid reward shaping, supervised warmup, and curriculum learning in enabling stable and sample-efficient RL training. By bridging RL with retrieval, SRAS opens a new direction for adaptive, task-aware, and lightweight RAG systems.

\noindent
Our main contributions are as follows:
\begin{itemize}
    \item We propose SRAS, a lightweight, reinforcement learning-based document selector that models query-document interactions and learns from sparse QA supervision.
    \item We introduce a hybrid reward signal combining relaxed F1 and BERTScore to guide learning without requiring dense annotations.
    \item We show that SRAS achieves strong QA performance under edge-device constraints, with sub-1MB model size and sub-1s inference latency.
    \item We provide detailed ablations highlighting the importance of reward shaping, supervised warmup, and curriculum learning for effective policy training.
    \item We evaluate SRAS zero-shot on 250 real-world QA examples from SQuAD v2 and demonstrate strong semantic generalization.
\end{itemize}

\section{Related Work}

\subsection{Static and Learned Retrieval in RAG Pipelines}
Retrieval-Augmented Generation (RAG) architectures augment language models with external documents retrieved from a corpus \cite{lewis2020retrieval, izacard2021leveraging}. Traditional approaches rely on static similarity-based retrieval using sparse methods like BM25 or dense methods such as Sentence-BERT \cite{reimers2019sentence}. These methods are efficient but lack adaptivity as they do not incorporate downstream task feedback or optimize retrieval for answer quality.

To address this, learned retrievers such as Dense Passage Retrieval (DPR) \cite{karpukhin2020dense} and ColBERT \cite{khattab2020colbert} have been proposed. These models are trained to embed queries and documents into a shared vector space using contrastive objectives. While they improve retrieval accuracy, their large size and compute requirements make them impractical for low-latency or edge scenarios.

\subsection{Reinforcement Learning for Retrieval and Ranking}
Reinforcement learning (RL) has been explored in information retrieval to optimize ranking policies from feedback signals \cite{wang2018reinforced, ai2019learning}. In QA systems, RL has been used to train retrievers using answer-level feedback rather than explicit document labels \cite{nogueira2019passage, xiong2021answer}. However, most prior work either trains full retriever encoders or combines retrieval and generation into a single RL agent, leading to high model complexity.

Recent work has explored modular training using reward signals derived from generation outputs to refine retrieval \cite{lu2021improving}. Nevertheless, these methods still rely on large pretrained encoders, making them unsuitable for constrained environments.

\subsection{Document Selection under Resource Constraints}
Deploying RAG systems on edge devices introduces unique challenges: memory footprint, inference latency, and power consumption must be minimized without sacrificing accuracy \cite{edgeai2021survey}. Most learned retrievers are ill-suited for such settings, requiring GPU acceleration and large memory buffers.

Lightweight selectors using shallow architectures or projection-based scoring functions have been proposed in the context of model compression and neural ranking \cite{zhang2022lite}. However, their adaptation to QA-driven RAG pipelines with sparse supervision remains under-explored. Moreover, few studies directly target RL-based document selection under edge constraints.

In contrast, our work introduces a compact feedforward selector that learns to choose relevant documents using sparse QA-derived rewards. By decoupling retrieval from heavy encoders and focusing on learning from weak supervision, SRAS offers a new paradigm for low-resource RAG.

\section{Methodology}

We propose a modular, reinforcement learning-based retrieval framework designed for edge-friendly RAG deployments. At the core of this framework is \textit{SRAS} (Sparse Reward-Aware Selector), a compact neural selector trained to optimize document selection using weak supervision derived from downstream QA performance. This section outlines the architecture of the pipeline, the reward formulation, and the training procedure using Proximal Policy Optimization (PPO) \cite{schulman2017ppo}.

\subsection{Pipeline Overview}

Our SRAS-enhanced RAG pipeline consists of four modular stages:

\begin{enumerate}
    \item \textbf{Corpus Preprocessing and Embedding:} All raw documents are preprocessed into a flat JSON structure and embedded using a lightweight Sentence-BERT encoder (MiniLM-L6-v2) \cite{reimers2019sentence}, with outputs cached in \texttt{.npy} and \texttt{.pt} formats for reuse.

    \item \textbf{Synthetic QA Pair Generation:} A generative model (\texttt{valhalla/t5-base-qg-hl}) generates synthetic question-answer pairs over the corpus. These serve as training samples for evaluating document selection effectiveness.

    \item \textbf{Document Selection via SRAS:} Given a question embedding and a candidate set of document embeddings (typically $n=8$), SRAS assigns scores via a lightweight feedforward scoring function that models interactions between query and documents, and selects the top-$k$ documents for downstream use.

    \item \textbf{Answer Generation and Reward Computation:} The selected documents are fed to a frozen QA model (e.g., T5 or FiD) to generate an answer. The generated output is compared with the gold answer to compute a hybrid reward signal. This reward is then used to update the SRAS policy.
\end{enumerate}

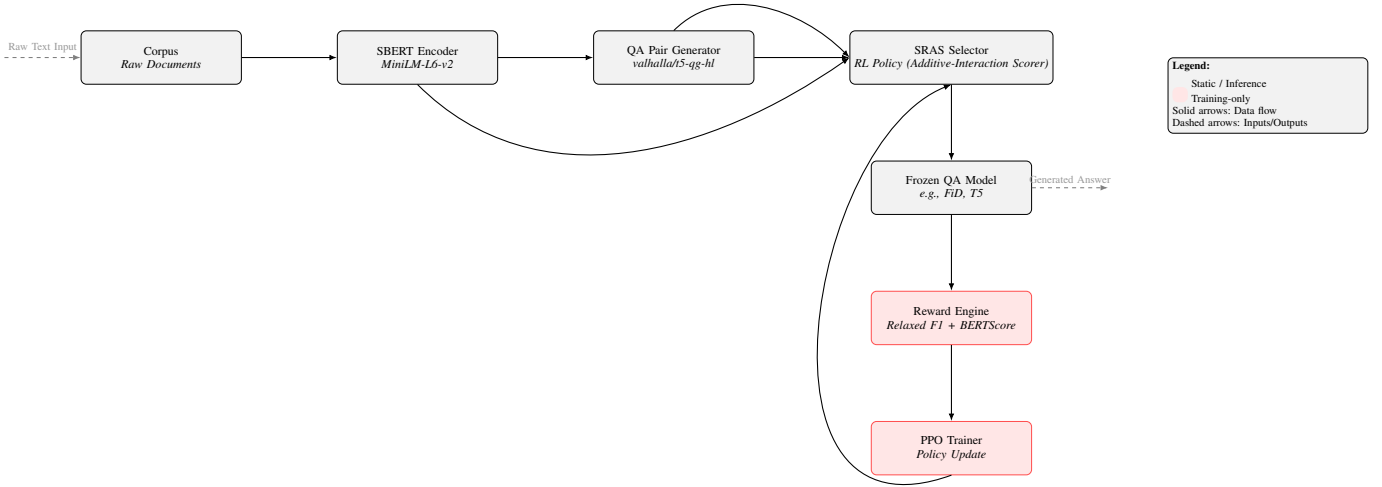
\begin{figure*}[htbp]
\centering
\begin{adjustbox}{width=\textwidth, trim=0pt 0pt 0pt 0pt}
\begin{tikzpicture}[
    node distance = 2cm and 2.5cm,
    module/.style = {
        rectangle, draw, rounded corners,
        minimum width=4.2cm, minimum height=1.4cm,
        align=center, font=\small, fill=gray!10
    },
    training/.style = {
        module, fill=red!10, draw=red!70
    },
    arrow/.style = {thick, -{Latex[length=2mm]}},
    ioarrow/.style = {thick, dashed, -{Latex[length=2mm]}, gray},
    labelstyle/.style = {font=\footnotesize, text=gray!80},
]

\node[module] (corpus) {Corpus\\\textit{Raw Documents}};
\node[module, right=of corpus] (sbert) {SBERT Encoder\\\textit{MiniLM-L6-v2}};
\node[module, right=of sbert] (qg) {QA Pair Generator\\\textit{valhalla/t5-qg-hl}};
\node[module, right=of qg] (selector) {SRAS Selector\\\textit{RL Policy (Additive-Interaction Scorer)}};

\node[module, below=of selector] (reader) {Frozen QA Model\\\textit{e.g., FiD, T5}};
\node[training, below=of reader] (reward) {Reward Engine\\\textit{Relaxed F1 + BERTScore}};
\node[training, below=of reward] (ppo) {PPO Trainer\\\textit{Policy Update}};

\draw[arrow] (corpus) -- (sbert);
\draw[arrow] (sbert) -- (qg);
\draw[arrow] (qg) -- (selector);
\draw[arrow] (selector) -- (reader);
\draw[arrow] (reader) -- (reward);
\draw[arrow] (reward) -- (ppo);

\draw[arrow] (sbert.south) to[out=-45, in=215] (selector.west);
\draw[arrow] (qg.north) to[out=45, in=135] (selector.west);

\draw[arrow] 
  (ppo.south) 
  .. controls +(-5.5,-2) and +(-3.75,-1.5) .. 
  (selector.south);

\draw[ioarrow] ([xshift=-2cm]corpus.west) -- (corpus.west) node[midway, above, labelstyle] {Raw Text Input};
\draw[ioarrow] (reader.east) -- ([xshift=2cm]reader.east) node[midway, above, labelstyle] {Generated Answer};

\node[rectangle, draw, rounded corners, fill=gray!10, font=\footnotesize, right=3cm of selector, anchor=north west, text width=5cm] (legend) {
    \textbf{Legend:} \\
    \tikz{\fill[gray!10] (0,0) rectangle (0.4,0.4);} Static / Inference \\
    \tikz{\fill[red!10] (0,0) rectangle (0.4,0.4);} Training-only \\
    Solid arrows: Data flow \\
    Dashed arrows: Inputs/Outputs
};

\end{tikzpicture}
\end{adjustbox}
\vspace{-3.5em}
\caption{End-to-end SRAS-enhanced RAG pipeline. The selector uses embeddings from both SBERT and the QA pair generator. Reward engine and PPO are used only during training.}
\label{fig:sras_rag_architecture}
\end{figure*}

Fig.~\ref{fig:sras_rag_architecture} illustrates the full architecture of the SRAS-enhanced RAG pipeline.

\subsection{Hybrid Reward Signal: QA-derived Feedback}

Since ground-truth labels for relevant documents are unavailable, we treat QA answer quality as a proxy for retrieval utility. Accordingly, we define a hybrid reward signal $R$ that combines both lexical and semantic similarity between the generated and reference answers:

\begin{equation}
    R = \alpha \cdot \text{Relaxed-F1} + (1 - \alpha) \cdot \text{BERTScore}
\end{equation}

We set $\alpha = 0.6$ based on a small-scale grid search over $\alpha \in \{0.3, 0.5, 0.6, 0.7\}$ using a held-out subset of the training QA pairs. Empirically, this value provided the best trade-off between early-stage lexical alignment (via Relaxed F1) and stable gradient signals (via BERTScore) for policy learning. Lower $\alpha$ values resulted in slower reward convergence, while higher values led to overfitting on token-level patterns.

\begin{itemize}
    \item \textbf{Relaxed F1:} A soft token-level F1 score computed over normalized answers (punctuation removed, lowercased, and stopwords removed), capturing partial lexical overlap and tolerance to minor phrasing differences.
    
    \item \textbf{BERTScore:} A semantic similarity metric that computes cosine similarity between contextualized token embeddings from a frozen RoBERTa-large model \cite{zhang2019bertscore}, capturing deeper meaning and paraphrase robustness.
\end{itemize}

This hybrid reward encourages SRAS to select documents that yield both lexically accurate and semantically faithful answers, while remaining robust to the weak supervision and redundancy present in synthetic QA pairs. The mixture also stabilizes PPO training by smoothing sparse or noisy reward signals during early exploration.

\subsection{SRAS Architecture}

The SRAS model is designed to be compact and hardware-efficient. Given a question embedding $q \in \mathbb{R}^d$ and a set of document embeddings $D = \{d_1, \ldots, d_n\} \subset \mathbb{R}^d$, we compute a relevance score $s_i$ for each document $d_i$ using:
\begin{align}
    h_q &= W_q q \in \mathbb{R}^h \\
    h_{d_i} &= W_d d_i \in \mathbb{R}^h \\
    s_i &= w^\top \tanh(h_q + h_{d_i}) \in \mathbb{R}
\end{align}
Where $W_q, W_d \in \mathbb{R}^{h \times d}$ are learnable projection matrices and $w \in \mathbb{R}^h$ is a learned attention vector. The model has approximately 197K parameters, resulting in a total size of $\sim$0.76 MB.

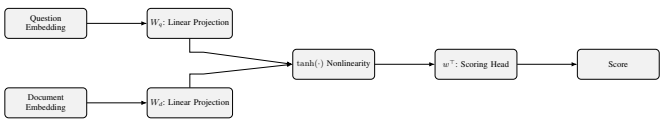
\begin{figure}[htbp]
\centering
\begin{adjustbox}{width=0.48\textwidth}
\begin{tikzpicture}[
    node distance = 1.8cm and 2.2cm,
    module/.style = {
        rectangle, draw, rounded corners,
        minimum width=3cm, minimum height=1.1cm,
        align=center, font=\small, fill=gray!10
    },
    arrow/.style = {thick, -{Latex[length=2mm]}},
    note/.style = {font=\footnotesize, text width=5cm, align=left}
]

\node[module] (q_embed) {Question\\Embedding};
\node[module, below=of q_embed] (d_embed) {Document\\Embedding};

\node[module, right=of q_embed] (q_proj) {$W_q$: Linear Projection};
\node[module, right=of d_embed] (d_proj) {$W_d$: Linear Projection};

\node[module, right=of q_proj, yshift=-1.5cm] (tanh) {$\tanh(\cdot)$ Nonlinearity};

\node[module, right=of tanh] (score) {$w^\top$: Scoring Head};
\node[module, right=of score] (score_out) {Score};

\draw[arrow] (q_embed) -- (q_proj);
\draw[arrow] (d_embed) -- (d_proj);
\draw[arrow] (q_proj.south) -- ++(0,-0.5) -- ++(0.5,0) -- (tanh.west);
\draw[arrow] (d_proj.north) -- ++(0,0.5) -- ++(0.5,0) -- (tanh.west);
\draw[arrow] (tanh) -- (score);
\draw[arrow] (score) -- (score_out);

\end{tikzpicture}
\end{adjustbox}
\caption{SRAS scoring architecture. Question and document embeddings are projected to a shared space, combined via a $\tanh$ nonlinearity, and scored linearly.}
\label{fig:sras_architecture}
\end{figure}

Fig.~\ref{fig:sras_architecture} presents a detailed breakdown of the SRAS scoring mechanism.

\subsection{Training with PPO under Sparse Rewards}

We adopt the PPO algorithm \cite{schulman2017ppo} to train SRAS from downstream QA-derived rewards. The document selection task is framed as a discrete action selection problem over $n$ candidates, where the agent selects $k$ documents per QA pair.

\subsubsection{PPO Configuration}

\begin{itemize}
    \item \textbf{Epochs:} 25
    \item \textbf{Batch Size:} 8
    \item \textbf{Top-$k$ Selection:} 3
    \item \textbf{Learning Rate:} $1 \times 10^{-5}$
    \item \textbf{Discount Factor $\gamma$:} 0.99
    \item \textbf{Clip $\epsilon$:} 0.2
    \item \textbf{Optimizer:} AdamW
\end{itemize}

\subsubsection{Stabilization Techniques}

To improve sample efficiency and robustness under sparse feedback, we employ:

\begin{itemize}
    \item \textbf{Supervised Warmup:} A cross-entropy pretraining phase using gold QA document labels to initialize the selector.
    \item \textbf{Reward Normalization:} Each batch is normalized to zero-mean and unit variance.
    \item \textbf{Advantage Estimation:} A GAE-like estimator is used to smooth advantage values.
    \item \textbf{Curriculum Learning:} We start training with easier QA pairs (higher top-1 overlap) and gradually increase difficulty.
\end{itemize}

These components jointly enable efficient PPO training under weak supervision while preserving low inference latency and memory usage.

\section{Experimental Setup}

This section outlines the dataset construction, candidate sampling, model baselines, evaluation protocol, and deployment constraints used to assess the SRAS selector under edge-suitable conditions.

\subsection{Synthetic QA Dataset}

To enable reward-driven document selection in the absence of gold labels, we construct a synthetic QA dataset from a curated corpus of 905 diverse documents. Each document is converted into a flat JSONL format for efficient indexing, embedding, and question-answer (QA) generation.

We employ the \texttt{valhalla/t5-base-qg-hl} (a pre-trained model) for generating QA pairs via a highlight-based prompting strategy \cite{mishra2021turq}. This results in a total of 750 high-quality QA pairs, each associated with one gold context document.

\subsection{Candidate Document Pooling}

For each QA pair, we construct a candidate pool of $n=8$ documents: 1 gold context and 7 distractor documents sampled randomly from the corpus. This setup emulates noisy retrieval settings while keeping selection complexity tractable for resource-constrained inference.

All documents and queries are encoded using a frozen MiniLM-L6-v2 Sentence-BERT encoder \cite{reimers2019sentence}. Embeddings are cached and reused across training and evaluation for efficiency.

\subsection{Evaluation Metrics}

We evaluate selector performance using both QA-centric and deployment-centric metrics:

\begin{itemize}
    \item \textbf{Relaxed F1:} Token-level F1 score computed after normalization (lowercasing, punctuation and stopword removal). This measures partial lexical overlap with ground truth answers.
    \item \textbf{BERTScore F1:} A semantic similarity score based on contextual embeddings from RoBERTa-large \cite{zhang2019bertscore}, capturing deeper alignment between generated and reference answers.
    \item \textbf{Latency:} Mean CPU inference time per query (ms), measured on Intel i5 CPU in single-threaded mode with batch size 1.
    \item \textbf{Model Size:} Serialized selector model size (MB), including weights and configuration.
\end{itemize}

\subsection{Deployment Constraints}

All selectors are evaluated in CPU-only environments to simulate real-world edge-device inference. The downstream QA model is frozen to T5-small, using greedy decoding (beam size 1) with a maximum output length of 32 tokens.

Inference time is measured with batch size 1 to reflect realistic per-query latency. All document embeddings and the QA model remain fixed across evaluations for fairness.

\subsection{Baselines}

We compare SRAS against the following standard and learned selectors, summarized in Table~\ref{tab:baselines}:

\begin{itemize}
    \item \textbf{Top-$k$ Cosine:} A static dense retriever that ranks documents by cosine similarity with the query embedding using MiniLM SBERT.
    \item \textbf{Random:} Uniformly samples 3 documents from the candidate set, providing a performance lower bound.
    \item \textbf{Supervised (FF):} A feedforward neural selector trained using cross-entropy loss on gold document labels. Matches SRAS architecture.
    \item \textbf{SRAS (PPO):} Our proposed selector, trained end-to-end using PPO with hybrid QA-based rewards and no access to document labels.
\end{itemize}

All models select $k=3$ documents from a candidate pool of 8, and output is passed to the same frozen QA model.

\begin{table}[ht]
\centering
\caption{Selector Baselines Compared}
\label{tab:baselines}
\begin{tabular}{|l|l|c|c|}
\hline
\textbf{Selector} & \textbf{Scoring Function} & \textbf{Train Method} & \textbf{Learned?} \\
\hline
Top-$k$ Cosine     & Cosine Sim. (SBERT)      & None       & No \\
Random             & Uniform Sampling         & None       & No \\
Supervised (FF)    & $\tanh(W_q q + W_d d_i)$ & Cross-Entropy & Yes \\
SRAS (PPO)         & $\tanh(W_q q + W_d d_i)$ & PPO         & Yes \\
\hline
\end{tabular}
\end{table}

\subsection{Design Choices for Prototyping:}
To maintain tractability during reinforcement learning (RL) training and ensure controlled evaluation across ablation variants, we deliberately restrict our corpus to 100 documents and employ a frozen QA generation pipeline. This constraint enables reproducible reward computation and fair benchmarking, particularly in the presence of sparse and delayed supervision. While such simplifications may limit real-world scale, they allow us to isolate and evaluate the impact of policy enhancements (e.g., reward shaping, curriculum learning) with minimal confounding. The lightweight setup further aligns with our goal of prototyping efficient on-device RAG selectors.
\section{Results and Discussion}

This section presents a comparative evaluation of SRAS against both non-learned and learned document selectors under edge-device constraints. We analyze answer quality, latency, and the contribution of training enhancements using multiple quantitative metrics and visualizations.

\subsection{Document Selection Performance}

Table~\ref{tab:selector_results} reports the performance of all document selectors on a held-out test set of 300 QA pairs. SRAS (PPO Base) achieves a Relaxed F1 of 0.1473 and a BERTScore F1 of 0.8463, outperforming the supervised selector on Relaxed F1 while closely matching it on semantic similarity.

Interestingly, the Top-$k$ cosine baseline performs best overall (Relaxed F1: 0.1604, BERTScore F1: 0.8549), but it is a static, non-trainable approach relying solely on frozen MiniLM embeddings. While it offers zero latency and model footprint, it lacks adaptability and does not leverage task-specific feedback.

\begin{table}[htbp]
\centering
\caption{Performance of Document Selectors Under Edge Constraints (Evaluated on 300 QA Pairs)}
\label{tab:selector_results}
\begin{adjustbox}{margin*= -1em 0em 0em 0em}
\begin{tabular}{|l|c|c|c|c|}
\hline
\textbf{Selector} & \textbf{Relaxed F1} & \textbf{BERTScore} & \textbf{Latency (s)} & \textbf{Size (MB)} \\
\hline
Top-$k$ Cosine     & 0.1604 & 0.8549 & 0.07  & 0.00 \\
Random             & 0.1182 & 0.8344 & 0.10  & 0.00 \\
Supervised (FF)    & 0.1323 & 0.8511 & 0.46  & 0.76 \\
SRAS (PPO)         & 0.1473 & 0.8463 & 0.38  & 0.76 \\
\hline
\end{tabular}
\end{adjustbox}
\end{table}

\begin{figure}[htbp]
\centering
\includegraphics[width=0.95\linewidth]{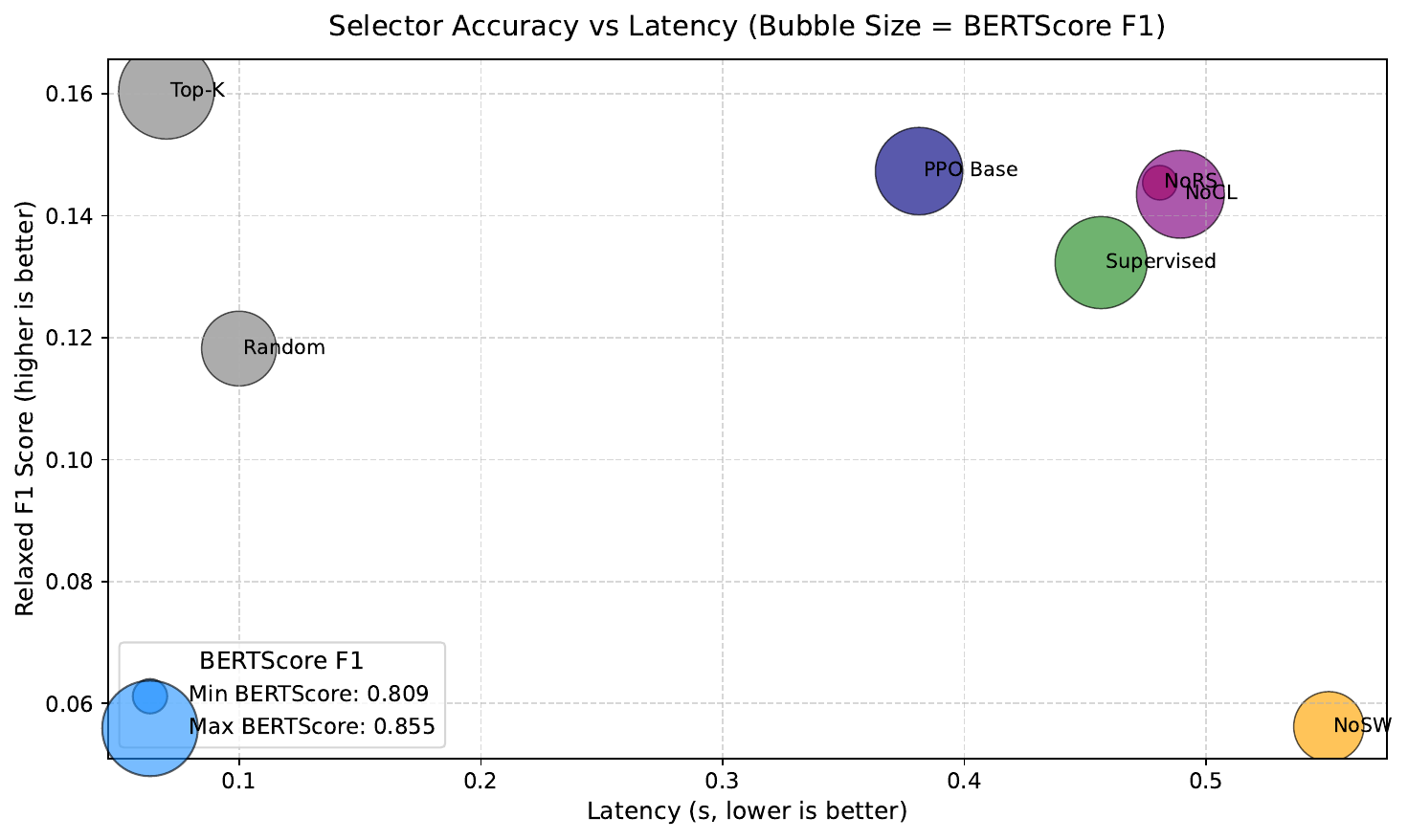}
\vspace{-1em}
\caption{SRAS achieves strong QA quality with low latency.}
\label{fig:latency_vs_f1}
\end{figure}

To better illustrate the trade-offs, Fig.~\ref{fig:latency_vs_f1} presents a latency vs. Relaxed F1 bubble plot, where bubble size denotes BERTScore F1. SRAS achieves a strong balance between answer quality and inference efficiency, offering near-supervised accuracy with sub-second latency and a compact model size.

\subsection{Impact of Reinforcement Learning Enhancements}

\begin{figure}[htbp]
\centering
\includegraphics[width=0.95\linewidth]{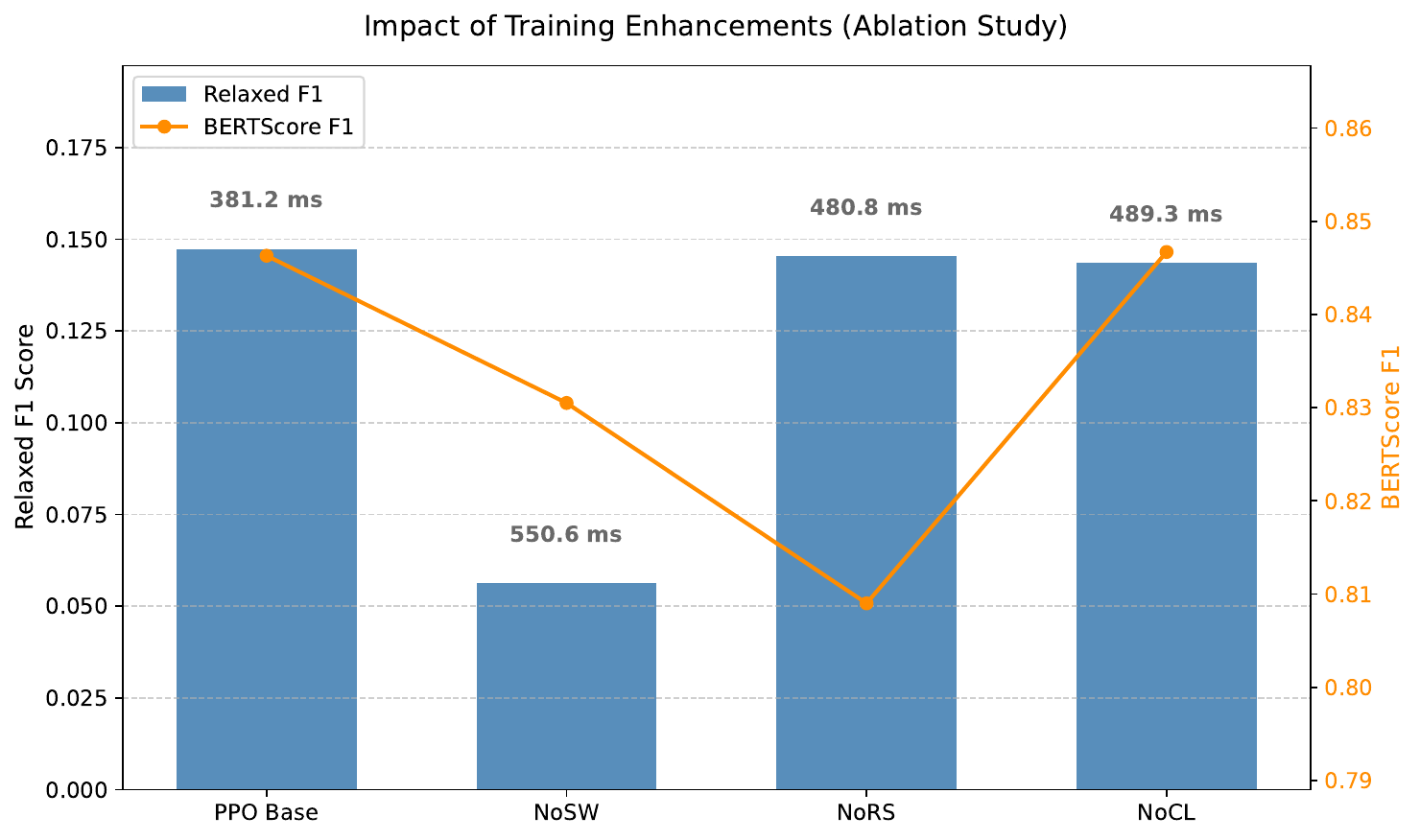}
\caption{Ablation study showing the effect of removing supervised warmup (SW), reward shaping (RS), and curriculum learning (CL).}
\label{fig:ablation_bar}
\end{figure}

To isolate the contribution of individual PPO training strategies, we perform an ablation study disabling: (i) supervised warmup (NoSW), (ii) reward shaping (NoRS), and (iii) curriculum learning (NoCL). Fig.~\ref{fig:ablation_bar} presents the Relaxed F1 and BERTScore F1 for each variant, with latency annotated on the bars.

\textbf{Key Findings:}
\begin{itemize}
    \item \textbf{Reward Shaping (RS)}: Its removal causes the steepest drop in Relaxed F1 (0.1473 $\rightarrow$ 0.0562), showing that sparse QA rewards alone are insufficient for effective learning.
    \item \textbf{Supervised Warmup (SW)}: NoSW results in undertraining (BERTScore F1 drops to 0.8305) and higher latency, underscoring the value of bootstrapping from labeled supervision.
    \item \textbf{Curriculum Learning (CL)}: NoCL slightly reduces Relaxed F1 (to 0.1435) and increases reward variance, suggesting it improves training stability more than final performance.
\end{itemize}

\subsection{PPO Reward Convergence}

\begin{figure}[ht]
\centering
\includegraphics[width=\linewidth]{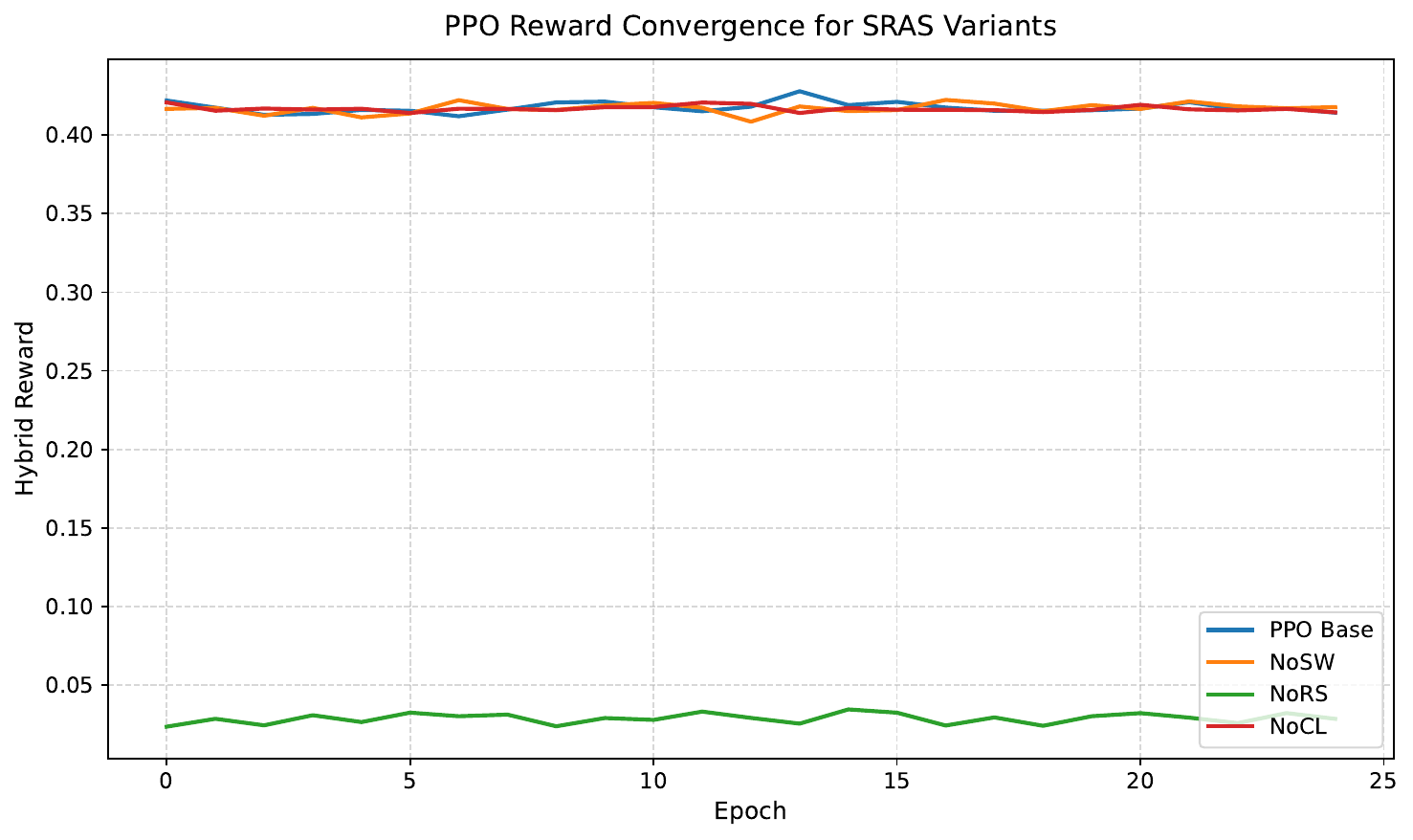}
\vspace{-1.5em}
\caption{Average PPO reward per epoch. Removing RS stalls learning. SW and CL improve early training stability.}
\label{fig:ppo_rewards}
\end{figure}

Fig.~\ref{fig:ppo_rewards} shows reward progression across training epochs for PPO Base and the ablation variants. The full PPO agent (with SW, RS, and CL) converges stably around a reward of 0.42.

\textbf{NoRS} fails to learn, plateauing around 0.02–0.03 reward, confirming the importance of shaped rewards in sparse-feedback settings.

\textbf{NoSW} achieves similar final rewards but exhibits greater volatility during early training, highlighting how supervised initialization improves stability.

\textbf{NoCL} converges comparably but with less consistent progression, indicating that curriculum learning improves training smoothness rather than outcome.

\subsection{Generalization to Real-World QA: SQuAD v2 Evaluation}

To evaluate SRAS under real-world conditions, we tested it on 250 QA pairs from SQuAD v2 - a benchmark featuring answerable and unanswerable questions. All selectors (SRAS, Top-$k$, Random) select one document per question, followed by answer generation using T5-base. Answers are evaluated against gold spans using Relaxed F1 and BERTScore F1.

\begin{table}[htbp]
\centering
\caption{Performance on 250 QA pairs from SQuAD v2.}
\label{tab:squad_eval}
\begin{tabular}{|l|c|c|}
\hline
\textbf{Selector} & \textbf{Relaxed F1} & \textbf{BERTScore F1} \\
\hline
SRAS (PPO Base) & \textbf{0.0454} & \textbf{0.8546} \\
Random          & 0.0313          & 0.0956          \\
Top-$k$ Cosine  & 0.0256          & 0.1282          \\
\hline
\end{tabular}
\end{table}

Despite being trained on synthetic data, SRAS generalizes effectively, achieving the highest semantic alignment (BERTScore F1: 0.8546) and outperforming both baselines.

\begin{figure}[htbp]
\centering
\includegraphics[width=0.9\linewidth]{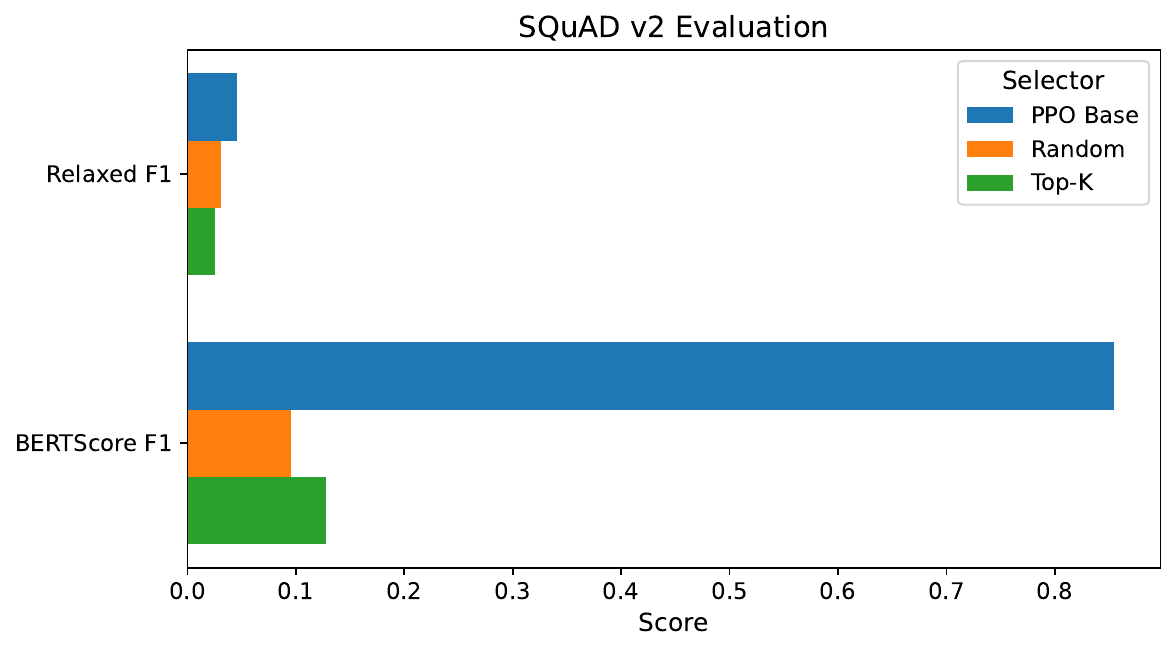}
\caption{SRAS vs. baselines on SQuAD v2: Relaxed F1 and BERTScore F1. SRAS generalizes well despite domain shift.}
\label{fig:squad_barplot}
\end{figure}

\section{Conclusion}

We proposed SRAS (Sparse Reward-Aware Selector), a lightweight reinforcement learning-based document selection framework for retrieval-augmented generation (RAG) pipelines under edge constraints. SRAS replaces conventional top-$k$ retrieval with a compact policy trained via Proximal Policy Optimization (PPO), guided by QA-based task rewards. Unlike similarity-based heuristics or static classifiers, SRAS adaptively selects documents that maximize downstream QA performance and generalizes well to unseen queries.

To address challenges from sparse and delayed rewards in realistic QA scenarios, we introduced three key training enhancements: supervised warmup, reward shaping, and curriculum learning. Ablations show each is essential: reward shaping speeds early learning, supervised warm-up stabilizes policy initialization, and curriculum learning enables scaling to larger candidate pools.

On a 300-example QA benchmark (using existing corpus data), SRAS achieves a Relaxed F1 of 0.1473 and BERTScore F1 of 0.8463, competitive with supervised baselines and near oracle top-$k$ retrieval, while maintaining sub-second latency and a model size under 1MB. Evaluation on 250 unseen SQuAD v2 questions confirms SRAS’s ability to generalize beyond its training distribution.

These traits make SRAS highly suitable for compute-constrained deployment.

\subsection*{Limitations and Future Work}

SRAS is trained on a static extractive QA corpus with limited domain diversity. Future work includes scaling to multi-domain and multilingual corpora, extending to generative or abstractive QA tasks, and integrating with differentiable retrievers for end-to-end training. We plan to evaluate SRAS on embedded hardware and explore further compression via quantization and pruning. Currently, the retriever is decoupled and fixed; integrating SRAS into end-to-end retriever-generator frameworks remains future work.

Ultimately, SRAS takes a step toward real-time, high-quality document selection in edge-native RAG systems: bridging the gap between learning-based retrieval and practical on-device intelligence.

\section*{Acknowledgment}
The authors acknowledge the use of OpenAI’s ChatGPT for language refinement and minor formatting support during manuscript preparation. All research concepts, methodology, experiments, and results are solely the work of the authors.

\end{document}